\documentstyle[12pt]{article}
\input tcilatex
\begin{document}

\author{D. Alves and J. F. Fontanari \\
Instituto de F\'\i sica de S\~ao Carlos - USP}
\title{A population genetics approach to the quasispecies model}
\date{}
\maketitle

\begin{abstract}
A population genetics formulation of Eigen's molecular quasispecies model
[Naturwissenchaften {\bf 58}, 465 (1971)] is proposed and several simple
replication landscapes are investigated analytically. Our results show a
remarkable similarity to those obtained with the original kinetics
formulation of the quasispecies model. However, due to the simplicity of our
approach, the space of the parameters that define the model can be
thoroughly explored. In particular, for the single-sharp-peak landscape our
analysis yields some interesting predictions such as the existence of a
maximum peak height and a minimum molecule length for the onset of the error
threshold transition.
\end{abstract}


\section{Introduction}

\label{sec:level1}

The molecular quasispecies model introduced by Manfred Eigen more than
twenty years ago \cite{Eigen} has become a major framework of the research
on the dynamics of competing self-reproducing macromolecules (see \cite
{reviews} for a review). In this model, a molecule is represented by a
string of $\nu $ digits $\left( s_1,s_2,\ldots ,s_\nu \right) $, with the
variables $s_i$ allowed to take on $\kappa $ different values, each of which
representing a different type of monomer used to build the molecule. The
number of different types of molecules is thus $\kappa ^\nu $. The
concentrations $x_i$ of molecules of type $i=1,2,\ldots ,\kappa ^\nu $
evolve in time according to the following differential equations 
\begin{equation}
\frac{dx_i}{dt}=\sum_jW_{ij}x_j-\left[ D_i+\Phi \left( t\right) \right]
x_i\;,  \label{ODE}
\end{equation}
where the constants $D_i$ stand for the death probability of molecules of
type $i$ and $\Phi (t)$ is a dilution flux that keeps the total
concentration constant. The elements of the replication matrix $W_{ij}$
depend on the replication rate $A_i$ of molecules of type $i$ as well as on
the Hamming distance $d\left( i,j\right) $ between strings $i$ and $j$. They
are given by 
\begin{equation}
W_{ii}=A_i\,q^\nu
\end{equation}
and 
\begin{equation}
W_{ij}=\frac{A_i}{\left( \kappa -1\right) ^{d\left( i,j\right) }}\,q^{\nu
-d\left( i,j\right) }\left( 1-q\right) ^{d\left( i,j\right) }~~~~i\neq j,
\end{equation}
where $0\leq q\leq 1$ is the single-digit replication accuracy, which is
assumed to be the same for all digits. Perhaps the main outcome of the
quasispecies model is that, for a given replication accuracy, there exists a
maximum string length which selection can maintain. This phenomenon, termed
the error threshold, poses a serious difficulty in envisioning life as an
emergent property of systems of competing self-replicating macromolecules.
It seems that some sort of cooperation between the macromolecules must be
incorporated in the model in order to avoid this error catastrophe \cite
{Hyper,Kauffman}.

In this paper we employ a classic population genetics approach \cite{Hartl}
to investigate the evolution of an infinite population of self-replicating
molecules. To accomplish that we have to make two simplifying assumptions to
the original quasispecies model. First, we assume that molecules composed of
the same number of monomers of each type are equivalent, i.e., possess the
same replication rate, regardless of the particular positions of the
monomers inside the molecules. Hence, a given molecule is characterized
solely by the vector $\vec{P}= \left ( P_1, P_2, \ldots, P_\kappa \right )$
where $P_\alpha$ is the number of monomers of type $\alpha$ in that
molecule. Since $\sum_\alpha^\kappa P_\alpha = \nu$, the number of different
types of molecules is reduced to $\left ( \nu + \kappa -1 \right ) !/ \nu !
\left ( \kappa -1 \right ) ! $. Second, in the population genetics approach
we focus on the evolution of the monomer frequencies, rather than on the
evolution of the molecule frequencies or concentrations. Henceforth the
variable $t$ will denote the number of nonoverlapping generations or simply
the generation number. We assume then that, given the monomer frequencies in
generation $t$, $p_\alpha (t) $ with $\sum_\alpha^\kappa p_\alpha (t)= 1$,
the molecule frequencies are given by the multinomial distribution 
\begin{equation}  \label{multinomial}
\Pi_t (\vec{P}) = C_{\vec{P}}^\nu \left [ p_1(t) \right ]^{P_1} \left [
p_2(t) \right ]^{P_2} \ldots \left [ p_\kappa(t) \right ]^{P_\kappa}
\end{equation}
where $C_{\vec{P}}^\nu = \nu !/P_1! P_2! \ldots P_\kappa!$. Thus, in each
generation the monomers are sampled with replacement from an infinite
monomer-pool. The effects of random drift are neglectable because the
population of molecules is also infinite. The changes in the monomer
frequencies are due then to the driving of natural selection, modelled by
the replication rate $A ( \vec{P} )$, and to mutations, modelled by the
error rate per digit $1-q$. A similar assumption was employed in the
analytical study of the effects of learning on evolution \cite{Fontanari}.
With these assumptions we are able to study analytically the dynamical
behavior of the model in the full space of the control parameters $\nu$, $q$%
, $\kappa$, and replication landscapes $A ( \vec{P} )$. In particular, while
previous investigations \cite{reviews} have almost exclusively dealt with
binary strings ($\kappa = 2$), our population genetics approach readily
applies to the analysis of more complex strings.

A worth mentioning result concerning the quasispecies model is the existence
of a correspondence between the ordinary differential equations (\ref{ODE})
and the equilibrium properties of a surface lattice systems \cite{Lethausser}%
. However, from an operational viewpoint, it seems easier to solve directly
the ordinary differential equations than to use cumbersome statistical
mechanics tools to obtain the surface equilibrium properties of the
corresponding lattice system for finite $\nu$ \cite{Tarazona}. Actually,
most of the statistical mechanics analyses of the quasispecies model are
restricted to the limit $\nu \rightarrow \infty$ \cite
{Lethausser,Tarazona,Peliti}.

The remainder of this paper is organized as follows. In Sec.\ \ref
{sec:level2} we derive the equations governing the evolution of the monomer
frequencies. To better appreciate the consequences of our simplifying
assumptions, these equations are solved for several simple replication
landscapes that have already been thoroughly analysed in the literature \cite
{reviews,Tarazona}. In Sec.\ \ref{sec:level3} we discuss our results and
present some concluding remarks. In particular, we point out how the model
can be generalized so as to include sexual reproduction between the
molecules.

\section{The model}

\label{sec:level2}

We now proceed with the derivation of the basic recursion relations for the
monomer frequencies. The fraction of monomers of type $\alpha $ that a
molecule characterized by $\vec P$ contributes to generation $t+1$ is
proportional to the product of three factors: (a) its frequency in the
population $\Pi _t(\vec P)$ in generation $t$, (b) its replication rate $%
A(\vec P)$, and (c) the average number of monomers $\alpha $ that replicate
correctly, $qP_\alpha $, plus the average number of monomers $\beta \neq
\alpha $ that due to replication errors mutate to $\alpha $, $(1-q)/(\kappa
-1)\sum_{\beta \neq \alpha }P_\beta $. After some simple algebra it yields 
\begin{equation}
p_\alpha (t+1)=\frac 1{\kappa -1}\,\left( 1-q+\frac{\kappa q-1}{w_t}%
\;\sum_{\vec P}\Pi _t(\vec P)\,A(\vec P)\,P_\alpha \right) ,  \label{freqgen}
\end{equation}
where the normalization factor $w_t$ is the average replication rate of the
entire population in generation $t$, 
\begin{equation}
w_t=\nu \sum_{\vec P}\Pi _t(\vec P)\,A(\vec P).
\end{equation}
Here the notation $\sum_{\vec P}$ stands for $\sum_{P_1=0}^\nu \ldots
\sum_{P_\kappa =0}^\nu \delta \left( \nu ,\sum_\alpha ^\kappa P_\alpha
\right) $, where $\delta (k,l)$ is the Kronecker delta. To proceed further
we must specify the replication rate $A(\vec P)$ of each molecule type,
i.e., specify the replication landscape.

\subsection{Single sharp maximum}

In this case we ascribe replication rate $a$ to the molecule composed of $%
\nu $ monomers of type $1$ and replication rate $1$ to all the remaining
molecules, i.e., $A(\vec P)=a$ if $\vec P=(\nu ,0,\ldots ,0)$ and $A(\vec
P)=1$ otherwise. This is the simplest and probably the most studied
replication landscape \cite{reviews}, because it clearly shows that although
the so-called master string $(\nu ,0,\ldots ,0)$ has no match its chance of
successfully taking over the population depends nontrivially on the values
of the control parameters as well as on the initial monomer frequencies $%
p_\alpha (0)$. Hence Eq.\ (\ref{freqgen}) reduces to 
\begin{equation}
p_1(t+1)=\frac 1{\kappa -1}\,\left[ 1-q+\left( \kappa q-1\right) \,\frac{%
p_1(t)+(a-1)\left[ p_1(t)\right] ^\nu }{1+(a-1)\left[ p_1(t)\right] ^\nu }%
\right]  \label{p1single}
\end{equation}
and 
\begin{equation}
p_\alpha (t+1)=\frac 1{\kappa -1}\,\left[ 1-q+\left( \kappa q-1\right) \,%
\frac{p_\alpha (t)}{1+(a-1)\left[ p_1(t)\right] ^\nu }\right] ~~~\alpha \neq
1.  \label{palphasingle}
\end{equation}
For simplicity, we keep the symmetry between monomers of type $\alpha \neq 1$
by setting their initial frequencies to $p_\alpha (0)=\left( 1-p_1(0)\right)
/\left( \kappa -1\right) $. Furthermore we bias the initial population
towards the master string by choosing $p_1(0)\approx 1$.

In Fig.\ \ref{fig1} we present the steady-state molecule frequencies,
obtained by solving the recursion relation (\ref{p1single}) for $\nu =10$, $%
a=50$ and $\kappa =2$, as a function of the error rate per digit $1-q$.
Three distinct regimes can be identified. First, the direct replication
regime (DR), that occurs for $1-q\leq 1-q_t\approx 0.241$, is characterized
by a molecular population composed of a cloud of mutants around the master
string, termed a {\it quasispecies}. In this regime there is a high
proportion of type 1 monomers, i.e., the fixed point is $p_1^{*}\approx 1$.
This fixed point disappears discontinuously at the error threshold $1-q_t$.
We note that, in contrast to the original quasispecies model, the error
threshold transition is discontinuous. Second, the stochastic replication
regime (SR), that occurs for $1-q>1-q_t$, is characterized by the fixed
point $p_1^{*}\approx 1/2$, which corresponds to an almost uniform
distribution of monomer types. Third, the complementary replication regime
(CR) that sets in when the replication error is so high ($1-q>0.86$) that
the monomers are almost certain to mutate, so that the population oscilates
between the quasispecies and its complement. This regime exists only for
binary strings, since only in this case the complement of a string is unique.

Some comments regarding the role of the initial monomer frequencies $%
p_\alpha (0)$ are in order. For $1-q<0.239$ the high $p_1$ fixed point is
the only stable fixed point of (\ref{p1single}). Above that value, a second
stable fixed point, $p_1^{*}\approx 1/2$, appears. These fixed points
compete, such that there is an all-or-none selection. The winner, however,
is not determined by the replication rate only, but also by the initial
monomer frequency $p_1(0)$. As mentioned above, the high $p_1$ fixed point
disappears at the error threshold transition. We will return to this issue
in the analysis of the competition between two sharp maxima.

The behavior pattern of the molecule frequencies for $\kappa > 2$ is
qualitatively similar to that discussed above: the DR and SR regimes are
characterized by the fixed points $p_1^* \approx 1$ and $p_1^* \approx
1/\kappa$, respectively, while the CR regime is absent.

In the following we will focus on the dependence of the error threshold $%
1-q_t$ on the control parameters. The fixed points $p_1(t+1)=p_1(t)=p_1^{*}$
of the recursion relation (\ref{p1single}) are the roots of $f\left(
p\right) =0$ where 
\begin{equation}
f\left( p\right) =\left( 1-q\right) \left[ \kappa p-1+\left( \kappa
-1\right) \left( a-1\right) p^\nu \right] -\left( a-1\right) \left(
1-p\right) p^\nu .  \label{f}
\end{equation}
Numerical analysis of this function indicates that the discontinuous
disappearance of the fixed point $p_1^{*}\approx 1$, that is the cause of
the error threshold phenomenon in our model, coincides with the appearance
of a double root of $f(p)$. Hence, the error threshold $1-q_t$ can be easily
determined by solving $f(p)=0$ and $df(p)/dp=0$ for $p$ and $q=q_t$
simultaneously. Eliminating the term $p^L$ of these two equations yields the
following quadractic equation for $p$, 
\begin{equation}
\kappa \,\nu \,p^2-\left[ 1+\nu +q\,\kappa \left( \nu -1\right) \right]
\,p+q\,\nu =0,  \label{f'}
\end{equation}
which possesses real roots either for $q\,\kappa \leq 1$ or $q\,\kappa \geq
\left[ (\nu +1)/(\nu -1)\right] ^2$. Only the latter is relevant for the
analysis of the error threshold, since this phenomenon occurs in the high
replication accuracy region. The disappearance of the high $p_1$ fixed point
is associated to the larger root of (\ref{f'}), while the smaller root is
related to the appearance of the uniform fixed point $p_1^{*}\approx
1/\kappa $. In order to avoid the error threshold discontinuous transition
we must set the control parameters so as to violate the second inequality.
In particular, for $\nu $ and $\kappa $ fixed, the discontinuous transition
line $q_t=q_t(a)$ terminates at the critical point 
\begin{equation}
q_c=\frac 1\kappa \,\left( \frac{\nu +1}{\nu -1}\right) ^2  \label{qc}
\end{equation}
\begin{equation}
a_c=1+\kappa ^\nu \,\left( \frac{\nu -1}{\nu +1}\right) ^\nu \,\frac{\left(
\kappa -1\right) \left( \nu -1\right) ^2-4\nu }{\left( \kappa -1\right)
\left( \nu ^2-1\right) },  \label{ac}
\end{equation}
which for large $\nu $ become $q_c\approx 1/\kappa $ and $a_c\approx \mbox{e}%
^{-2}\kappa ^\nu $, respectively. We note that the critical point
coordinates $p_c$, $q_c$ and $a_c$ are obtained by solving the three
equations $f(p)=0$, $df(p)/dp=0$ and $d^2f(p)/dp^2=0$ simultaneously. The
condition $q_c\leq 1$ implies that there is a minimum string length, 
\begin{equation}
\nu _{min}=\frac{\sqrt{\kappa }+1}{\sqrt{\kappa }-1},  \label{Lmin}
\end{equation}
below which the error catastrophe does not occcur. In Fig.\ \ref{fig2} we
present the phase diagram in the space $(1-q,a)$ for $\nu =10$ and $\kappa
=2 $. The discontinuous transition between the phases DR and SR ends at the
critical point $1-q_c=0.251$ and $a_c=58.01$, while the transition between
the phases SR and CR seems to never disappear. Thus, for a given value of $%
\nu $ it is always possible to choose a sufficiently large value of $a$ so
that the phases CR and SR are no longer distinguishable. To the best of our
knowledge, there is no similar result reported for the original quasispecies
model. It must be noted, however, that due to the numerical difficulty of
solving the set of $\kappa ^\nu $ ordinary differential equations (\ref{ODE}%
), the space of parameters has not been adequately explored for that model.
In fact, the computational effort needed to study the evolution of a
population of molecules composed of more than two types of monomers ($\kappa
>2$) is so large that the important problem of the dependence of the error
threshold $1-q_t$ on the number of monomer types $\kappa $ has remained
unaddressed so far. In the population genetics framework, however, the
number of recursion relations increases linearly with $\kappa $, so this
parameter does not introduce any particular difficulty to our analysis.
Moreover, for the replication landscapes considered in this paper, in which
the replication rates of the molecules are determined by one type of monomer
only, the problem is reduced to the solution of a single recursion relation.
In Fig.\ \ref{fig3} we present the dependence of $1-q_t$ on $\kappa $ for $%
a=10$ and several values of $\nu $. Note that beyond $\kappa \approx 4$ the
error threshold is almost insensitive to further increase of $\kappa $.
Hence, in order to maximize the information content of the quasispecies, it
is advantageous to choose $\kappa $ as large as possible. Different values
of $a$ do not produce any qualitative change in this figure.

\subsection{Single smooth maximum}

In what follows we will consider the case $\kappa =2$ only. We assume that
the replication rates of the molecules increase with the number of monomers
of type 1 they possess, irrespective of the other monomer types. More
specifically, 
\begin{equation}
A(P_1)=1+\left( a-1\right) \,\left( \frac{P_1-\mu }{\nu -\mu }\right) ^\gamma
\label{smooth}
\end{equation}
if $P_1\geq \mu $ and $A(P_1)=1$ otherwise. Here $\mu $ is an integer that
can take on the values $0,1,\ldots ,\nu -1$, and $\gamma $ is a real,
positive variable. Clearly, $\mu $ measures the size of the flat region of
the replication landscape, while the parameter $\gamma $ determines the
smoothness of the landscape near $\mu $: the larger $\gamma $, the smoother
the landscape. The same procedure employed in the analysis of the
single-sharp maximum, which is recovered for $\mu =\nu -1$, can be used to
investigate the error threshold transition for the smooth replication
landscape (\ref{smooth}). In particular, in Fig.\ \ref{fig4} we show the
error rate per digit at the threshold transition, $1-q_t$, as a function of
the exponent $\gamma $ for $\nu =20$, $a=10$ and several values of $\mu $.
For a given $\mu $ there is a critical value $\gamma _c$ below which the
error threshold phenomenon does not occur. This exponent is shown in Fig.\ 
\ref{fig5} as a function of the ratio $\mu /\nu $. It is clear from these
figures that broad maxima (small $\mu $) can resist longer to the error
catastrophe or even avoid it depending on the value of the exponent $\gamma $%
. Large values of $\gamma $ actually increase the size of the flat region
and so they favour the appearance of the error threshold. Our results are in
agreement with a comment by Tarazona \cite{Tarazona} that the exponent with
which the replication landscape goes flat is germane to the onset of the
error threshold transition. Different values of $\nu $ and $a$ do not change
qualitatively these results.

\subsection{Two sharp maxima}

As before, we assume that $A(\vec P)$ depends on $P_1$ only. In this case
the replication landscape consists of two sharp maxima $A(P_1=0)=A_0$, $%
A(P_1=\nu )=A_\nu $, and $A(P_1)=1$ otherwise. In order to illustrate the
role of the initial monomer frequencies we present in Figs.\ \ref{fig6}, \ref
{fig7} and \ref{fig8} the frequency of type 1 monomers as a function of the
generation number $t$ for $\nu =20$, $\kappa =2$, $A_0=200$, $A_{20}=10$ and
several initial frequencies. The evolution for $1-q=0$ is shown in Fig.\ \ref
{fig6}. There are only two stable fixed points, namely, $p_1^{*}=1$ and $%
p_1^{*}=0$. Despite the large difference between the replication rates of
the molecules associated to these fixed points, their basins of attraction
are practically of the same size. They would be strictly equal if $%
A_0=A_{20} $. The main effect of a large replication rate in this case is to
speed up the convergence to the low $p_1$ fixed point. By increasing the
error rate a new stable fixed point $p_1^{*}\approx 1/2$, associated to the
stochastic replication regime, appears. The interplay of the three stable
fixed points is shown in Fig.\ \ref{fig7} for $1-q=0.01$. For nonzero
replication error rates, the basin of attraction of the low $p_1$ fixed
point is considerably larger than that of the high $p_1$ fixed point. Of
course, their basins of attraction have actually decreased as compared with
the case $1-q=0$. We note that the two quasispecies do no coexist: for a
given initial population there is an all-or-none selection. Finally, in
Fig.\ \ref{fig8} we present the evolution for $1-q=0.06$. The high $p_1$
fixed point, associated to the molecule with the smaller replication rate,
has disappeared and the stochastic replication fixed point has taken over
its basin of attraction. Further increase of the error rate $1-q$ will
eventually lead to the disappearance of the low $p_1$ fixed point too.

Within this framework we can easily study the competition between a sharp
maximum and a broad or smooth maximum \cite{Schuster&Swetina}. The results
show the same qualitative features as those presented above. In particular,
since the broader maximum possesses the larger error threshold $1-q_t$ (see
Fig.\ \ref{fig4} ) it plays the same role as the larger replication rate
maximum. We note that, in contrast to the original quasispecies model, there
is no selection transition in our model \cite{Schuster&Swetina}, which would
amount to a discontinuous transition between the low and the high $p_1$
fixed points, i.e., the former should take over the basin of attraction of
the latter.

\section{Discussion}

\label{sec:level3}

An interesting extension of the quasispecies model is the possibility of two
molecules exchanging matter during a collision. Clearly, the analogous to
this phenomenon in the population genetics approach is sexual reproduction.
More specifically, the collision (mating) between the molecules (parents) $%
\left( s_1^f,\ldots ,s_\nu ^f\right) $ and $\left( s_1^m,\ldots ,s_\nu
^m\right) $ produces the new molecules (offspring) $\left( s_1^f,\ldots
,s_{c-1}^f,s_c^m,\ldots ,s_\nu ^m\right) $ and $\left( s_1^m,\ldots
,s_{c-1}^m,s_c^f,\ldots ,s_\nu ^f\right) $, where the digit $0\leq c\leq \nu 
$ is the so-called crossover point. The number of offspring depends, of
course, on the replication rate of the parent molecules. Using the
assumptions presented in Sec.\ \ref{sec:level1}, it is straighforward to
derive the following recursion relations for the evolution of the monomer
frequencies: 
\begin{equation}
p_\alpha (t+1)=\frac 1{\kappa -1}\left[ 1-q+\frac{\kappa q-1}{2\,w_t}%
\sum_{\vec P^f}\sum_{\vec P^m}\Pi _t(\vec P^f,\vec P^m)A(\vec P^f,\vec
P^m)(P_\alpha ^f+P_\alpha ^m)\right] ,  \label{sex}
\end{equation}
where 
\begin{equation}
w_t=\nu \sum_{\vec P^f}\sum_{\vec P^m}\Pi _t(\vec P^f,\vec P^m)\,A(\vec
P^f,\vec P^m)
\end{equation}
is the average replication rate of the entire population and 
\begin{equation}
\Pi _t(\vec P^f,\vec P^m)=\Pi _t(\vec P^f)\Pi _t(\vec P^m)
\end{equation}
is the frequency of the collisions or matings between the molecules $\vec
P^f $ and $\vec P^m$. Here $A(\vec P^f,\vec P^m)$ determines the number of
offspring generated by the mating between these two molecules. As expected
since the positions of the monomers inside the molecules play no role in our
population genetics approach, the basic recursion relations (\ref{sex}) are
independent of the crossover point $c$. It is interesting to note that this
equation reduces to eq.\ (\ref{freqgen}) in the case that $A(\vec P^f,\vec
P^m)=A(\vec P^f)A(\vec P^m)$ and so the two reproduction modes, asexual and
sexual, yield the same results. We have investigated the steady-state
solutions of (\ref{sex}) under a variety of conditions, but found no
noteworthy difference from the previously presented results. We only mention
that by penalizing matings within a same class, i.e., $A(\vec P^f,\vec
P^m)=1 $ if $\vec P^f=\vec P^m$ we can obtain the formation of a
quasispecies (a master string surrounded by a cloud of mutants) even in the
regime of perfect replication accuracy $q=1$.

The critical, though natural, assumption of the population genetics approach
proposed in this paper is the use of the multinomial distribution (\ref
{multinomial}) for the molecule frequencies. Since this is a single-peaked
distribution, the coexistence of two or more quasispecies, which could only
be described by a multi-peaked distribution, is prevented {\it apriori}. In
the original quasispecies model such a coexistence is possible only in the
case of degenerate quasispecies. This is an important issue, since it would
be highly desirable to study the spontaneous formation of hypercycles within
the framework of the quasispecies model \cite{Hyper,Epstein}.

In this paper we have presented a population genetics formulation of the
classic quasispecies model proposed by Eigen \cite{Eigen}. Owing to its
extreme simplicity, this formulation may be useful, in the sense of having
the value of an approximation, to tackle problems for which the numerical
difficulty of solving the ordinary differential equations (\ref{ODE}) or
employing the statistical mechanics approach \cite{Lethausser} makes the
analysis prohibitive. Furthermore, even for the well-studied replication
landscape that consists of a single sharp peak, our population genetics
analysis has yielded some interesting and unexpected results such as the
existence of a maximum peak height (\ref{ac}) and a minimum string length (%
\ref{Lmin}) for the onset of the error catastrophe. It would be interesting
to investigate whether similar bounds exist for the original quasispecies
model.

I. R. Epstein, J.\ Theor.\ Biol.\ {\bf 78}, 271 (1979).

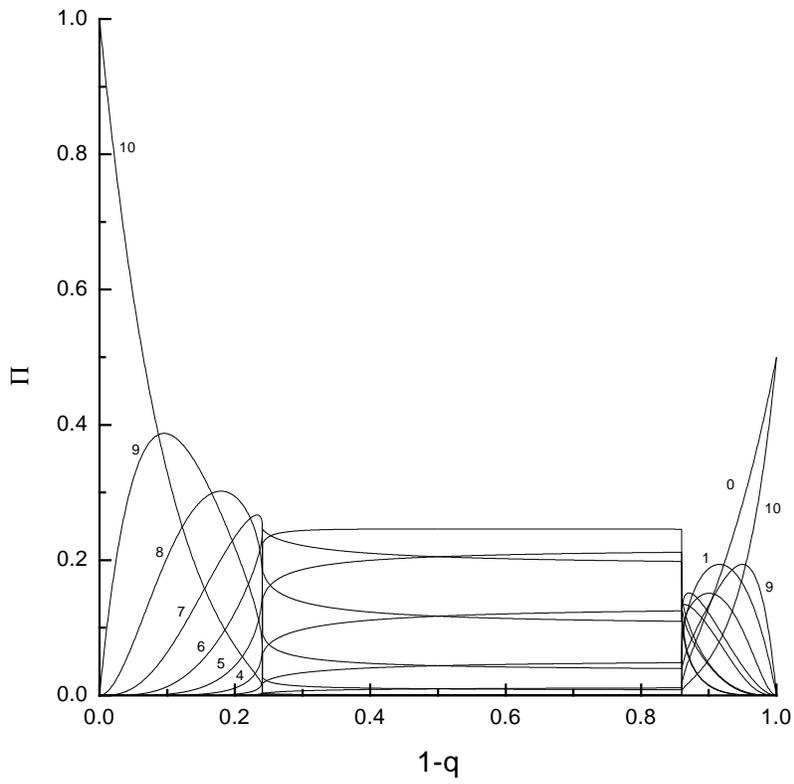
\begin{figure}[h]
\caption
{Steady-state frequencies of molecules of
type $P_1=10$ (master string), $9$, $8$, $7$, $6$, $5$, $4$ and $0$ as a
function of the error rate per digit $1-q$ for $\nu =10$, $a=50$ and $\kappa
=2$. The error threshold transition occurs at $1-q_t\approx 0.241$ and the
complementary replication regime sets in for $1-q>0.86$. }
\label{fig1}
\end{figure}
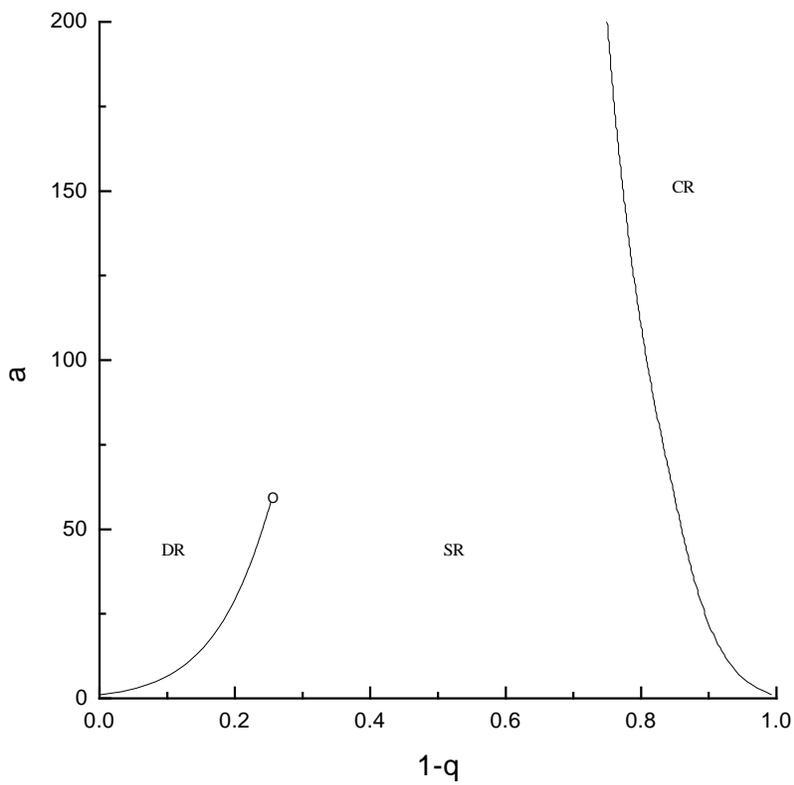
\begin{figure}[h]
\caption{Phase diagram in the space $(1-q,a)$ for $%
\nu =10$ and $\kappa =2$. The discontinuous transition between the phases DR
and SR ends at the critical point $1-q_c=0.251$ and $a_c=58.01$. }
\label{fig2}
\end{figure}
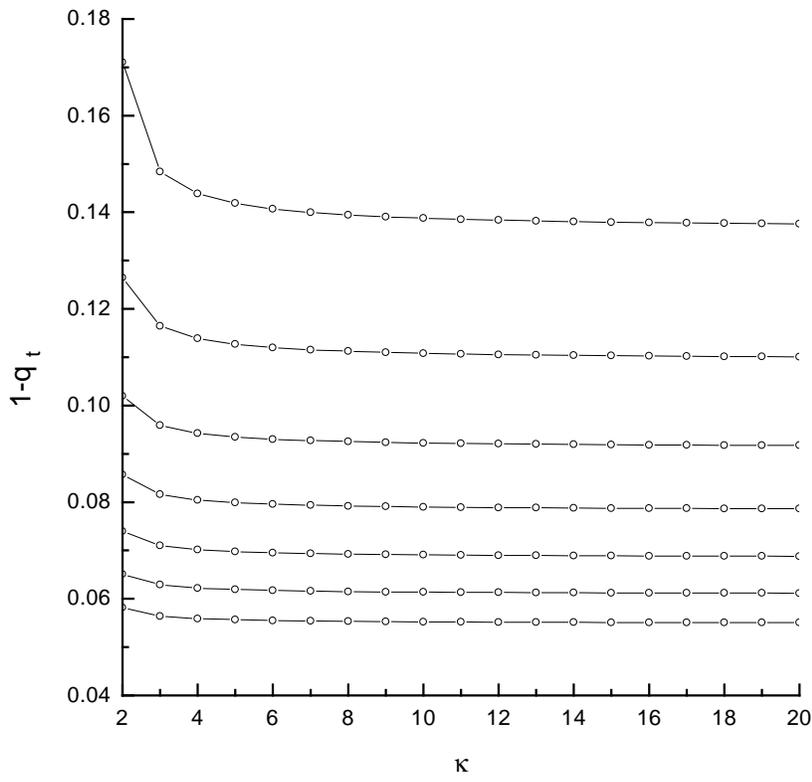
\begin{figure}[h]
\caption{Error threshold $1-q_t$ as a function of
the number of monomer types $\kappa $ for $a=10$ and (from top to bottom) $%
\nu =8$, $10$, $12$, $14$, $16$, $18$ and $20$. The lines are guides to the
eyes. }
\label{fig3}
\end{figure}
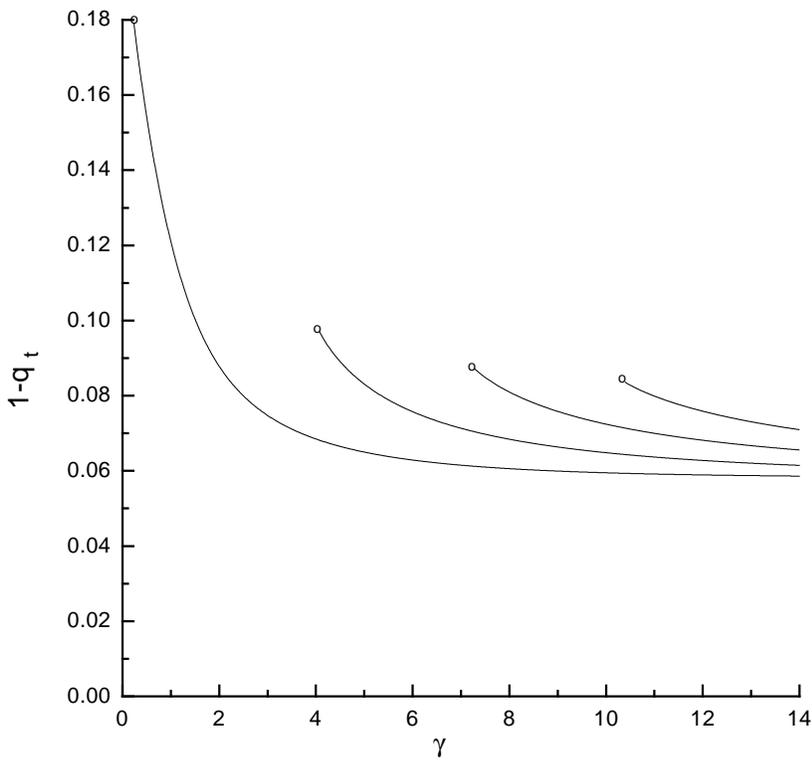
\begin{figure}[h]
\caption{Error threshold $1-q_t$ as a function of the exponent $\gamma$
for $\nu = 20$, $a = 10$ and (from left to right) $\mu/\nu = 0.75$, $0.5$, $%
0.25$ and $0$. The curves begin at $\gamma_c = \gamma_c ( \mu )$. }
\label{fig4}
\end{figure}
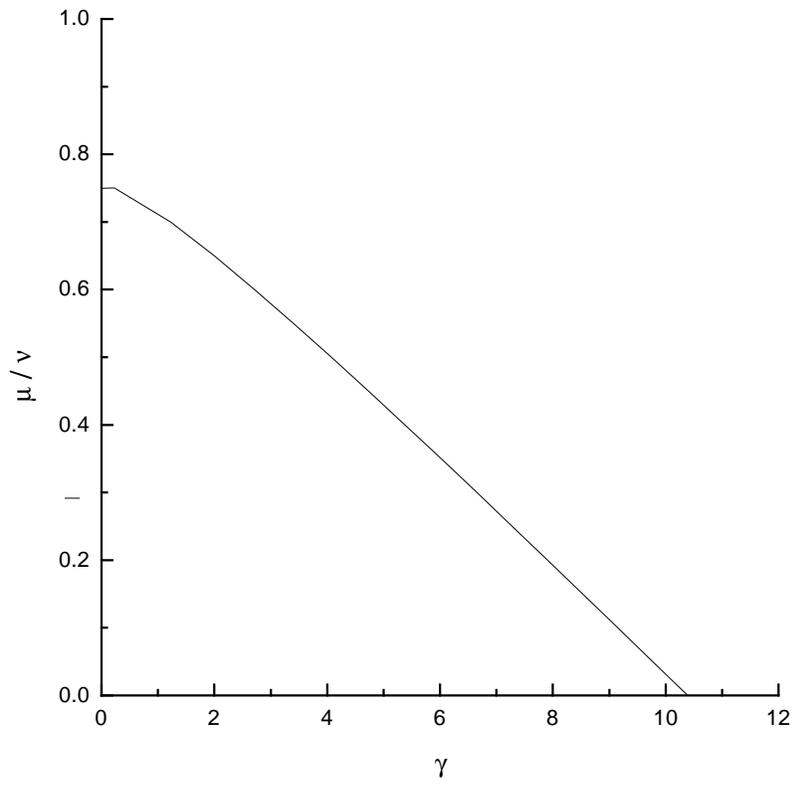
\begin{figure}[h]
\caption{Critical value of the exponent $\gamma$ as a function of the
ratio $\mu/\nu$ for $\nu = 20$ and $a = 10$. Below the curve the error
threshold phenomenon does not occur. }
\label{fig5}
\end{figure}
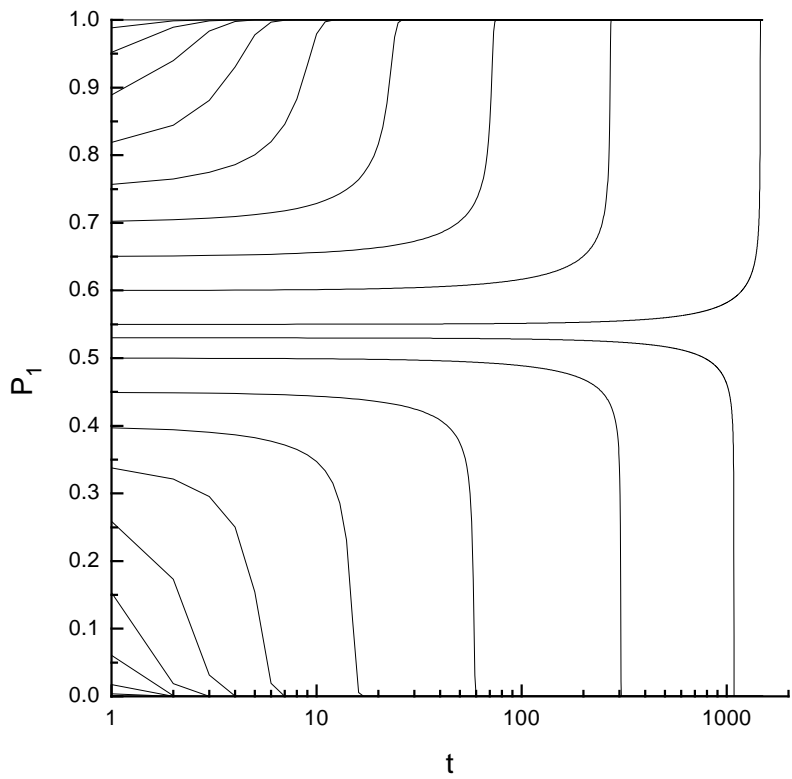
\begin{figure}[h]
\caption{Frequency of monomers of type 1 as a function of the generation
number for the two-sharp-maxima replication landscape and several initial
frequencies $p_1(0)$. The parameters are $\nu = 20$, $\kappa = 2$, $A_0 =
200 $, $A_{20} = 10$ and $1 - q = 0$. }
\label{fig6}
\end{figure}
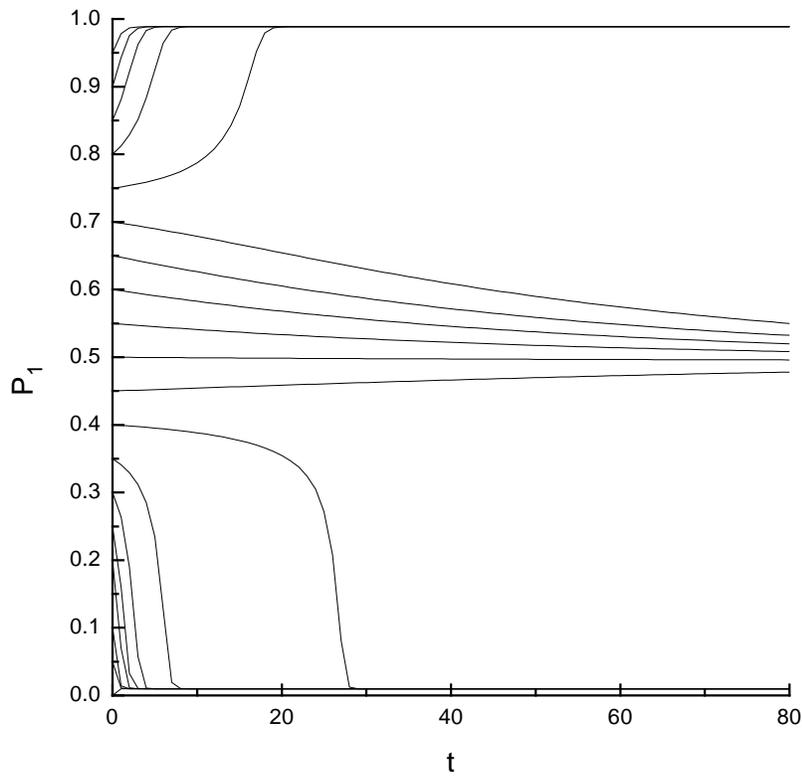
\begin{figure}[h]
\caption{Same as Fig. 6, but for $1 - q = 0.01$. }
\label{fig7}
\end{figure}
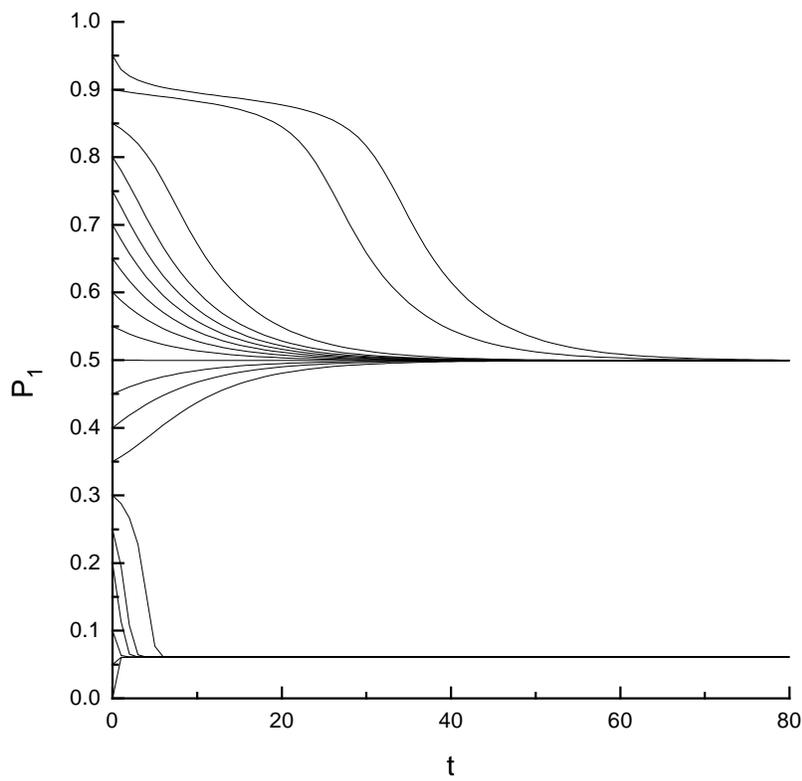
\begin{figure}[h]
\caption{Same as Fig. 6, but for $1 - q = 0.06$. }
\label{fig8}
\end{figure}

\end{document}